\documentclass[10pt,journal,letterpaper,compsoc]{IEEEtran}

\ifCLASSOPTIONcompsoc
\else
\fi
\ifCLASSINFOpdf
\else
\fi

\usepackage{amssymb,amsmath}
\usepackage{graphicx}
\usepackage{subfigure}
\usepackage{multicol}
\usepackage{multirow}
\usepackage{rotating}
\usepackage{algorithmic}
\usepackage{algorithm}
 \usepackage{amsthm}

\begin{document}
%

\title{A Comprehensive Survey of Watermarking Relational Databases
Research}

\author{M.~Kamran and \IEEEmembership{}Muddassar~Farooq~\IEEEmembership{}
\IEEEcompsocitemizethanks{\IEEEcompsocthanksitem M. Kamran is with
the Department of Computer Science, COMSATS Institute of Information Technology Wah Campus, Wah Cantt,
Pakistan.\protect\\
E-mail: muhammad.kamran@ciitwah.edu.pk \IEEEcompsocthanksitem Muddassar
Farooq is with the Muslim Youth University(MYU), Islamabad, Pakistan.
\protect\\E-mail: muddassar.farooq@gmail.com }
\thanks{}}
\markboth{}
{Shell \MakeLowercase{\textit{et al.}}: Bare Demo of IEEEtran.cls
for Computer Society Journals}

\IEEEcompsoctitleabstractindextext{%
\begin{abstract}
Watermarking and fingerprinting of relational databases are quite proficient for ownership protection, tamper proofing, and
proving data integrity. In past few years several such techniques have been proposed. A survey of almost all the work done, till date, in these fields has been presented in this paper. The techniques have been classified on the basis of how and where they embed the watermark. The analysis and comparison of these techniques on different merits has also been provided. In the end, this paper points out the direction of future research in these fields.
\end{abstract}

}

\maketitle

\IEEEdisplaynotcompsoctitleabstractindextext

\IEEEpeerreviewmaketitle

\section{Introduction}
\label{sec:Introduction}
The data is an important asset for its
owner. Digital data sharing is becoming an emerging trend both at the personal and organization level. Information and
communication technology (ICT) systems generate and use
enormous amount of data that contains useful knowledge which needs to be extracted by
using different data mining and warehousing techniques. In a collaborative environment, individuals and organizations need to
share their data using different resources.
The shared data must adhere to secrecy (for preventing unauthorized disclosure of data), integrity (malicious data modifications), and
availability (error recovery) \cite{bertino2005database}.
The digital data can be
copied, altered and may be re-distributed for different purposes,
and this fact may violate the copyrights of the data owner.

With an ever increase in sharing of data, the digital technology faces the challenge
of preventing piracy of precious and sensitive data in different formats like text, software, image, video, audio, spatial trajectory data
\cite{agrawal2003watermarking}, \cite{sion2007rights}, \cite{palsberg2000experience},
\cite{atallah2003natural}, \cite{jin2005watermarking} to ensure that the data is used in an authorized manner only.
Only the relevant and necessary data should be shared across private databases  and the shared data should also be
right protected.
To tackle this situation, there is a need of
an efficient right protection mechanism for the digital data. In this context, issues like ownership protection and illegitimate circulation
of digital content can be controlled by Digital Rights Protection (DRP).

Cryptographic techniques usually do not associate the cryptographic information with
digital contents; therefore, cryptography
cannot provide the ownership information \cite{yan2001mobile}. Moreover, the cryptosystem (a suite consisting of three algorithms: (i) a key generation algorithm; (ii) an encryption algorithm; and (iii) a decryption algorithm) does not provide any guaranteed mechanism for tracing the redistribution and/or alteration of the digital content.  Consequently, they alone are not well suited for DRP.
Some techniques such as \cite{goodrich2005indexing} can be used for detecting and identifying the changes made in the data -- by
organizing the indexing structure of stored data. Digital watermarking, on the other hand, associates permanent information
with the digital contents. This information can then be used to prove the ownership of the digital content. Watermarking
of digital data has been quite widely used for right protection of digital intellectual property (IP) \cite{abdel2004survey}.
A watermark is basically a piece of secret information, which is inserted into a specific location inside the digital
data as a copyright information.

Watermarking has -- so far -- found its application in right protection of
images, multimedia, natural language text, softwares, semi-structured data like XML, map information, relational databases,
electronic medical records (EMR), and Geographical Information Systems (GIS) databases.
Among these areas, relational databases watermarking
is a relatively fresh area but is of great importance because outsourced databases need to be right protected to stop its illegal (or unauthorized) use.
The watermarking of relational databases is different from other data formats. As opposed to other data formats,
in database watermarking one has to deal with multiple
tuples simultaneously, the ordering of database tuples does not matter, and a subset of the database
is also useable. Moreover, a database may contain attributes with different data types like numeric, text, and discrete data. So watermarking such data requires valid marks for every data type.

A database watermarking scheme should be
robust against all kinds of malicious attacks that are launched for deteriorating the
watermark. The watermarking scheme should also be imperceptible, which requires that there should not be any apparent difference between the original and the watermarked content. Furthermore, the watermark should be
such that it can be detected using some secret parameters like
secret key etc. A good watermarking technique is also blind, that is,
it should not need the original data or watermark to detect the
watermark from the altered data. Moreover, the embedded watermark should only
be detectable by the data owner. Furthermore, the watermark should not
deteriorate the original data and data usability should be ensured during the process of watermark insertion.
The watermarking of databases should also take
into account such methods which ensure that the query results and associated statistics should be preserved.
If a watermark has to be
embedded multiple times, then the watermarks should not affect each
other.

A fragile watermark, which is affected even by minor malicious attacks, is also a requirement for tamper proofing and data integrity. Fragile watermarking techniques are suitable for applications that require data integrity and therefore fragile watermarks are usually be vulnerable to even benign updates because he attacker --Mallory-- does not want to disturb the embedded watermark while trying to tamper the watermarked dataset.

%
%

Fingerprinting of relational databases is also similar to watermarking with a major difference: in fingerprinting
a database is marked with a different watermark for every different user of the fingerprinted database, while in watermarking usually the same
watermark is used for all the users of the watermarked database. One such technique is presented in \cite{kumar2013unauthorized}. The fingerprinting techniques are used in the applications domains where the data owner --Alice-- wants to identify the guilty agent who was responsible for the data leakage(the guilty agent is represented as Mallory in such scenarios.).

To summarize, the three main type of watermark techniques have different applications as:
\begin{itemize}
\item The robust watermarking techniques are suitable for applications that require ownership protection.
\item The fragile watermarking techniques serve the purpose for data tamperproofing and data integrity check.
\item The fingerprinting techniques provide a mechanism for identifying the guilty user whose data was compromised.
\end{itemize}

Arguably, the major requirement of watermarking scheme -- for ownership protection and fingerprinting -- is its robustness against malicious attacks.
Therefore, the data owner
Alice\footnote{Throughout this document, we refer Alice(female) as the data owner.},
would want to insert the most possible robust watermark in the database. Alice has two options to embed
watermark into her data. She can either modify the value of some selected bits of the data, or embed watermark
into the data statistics. The later approach is more robust to malicious attacks as the former can easily be attacked by simple
data manipulations, such as shifting of some least significant bits (LSBs). Also, the embedding of watermark in large number of
tuples makes it more resilient. But Alice can only insert the watermark bounded by the data
usability constraints. These usability constraints are application dependent, hence every time Alice wants to watermark her
data, she has to define the usability constraints.
Once Alice has watermarked her data, she would make it public or share (or sale) it
to some party. The attacker,
Mallory\footnote{Throughout this document, we refer Mallory (male) as the malicious attacker.}, would want to corrupt the
embedded watermark or even remove the watermark from the watermarked data. Since the attacker has no
access to the original database; therefore, he might deteriorate the usability of the data during these attacks. But it is supposed that he can
locate the watermark if he has access to the secret parameters (such as a \emph{secret key}). The attacker can perform every type of
manipulation on the watermarked data because he has unrestricted access to the watermarked data.
He can attack the data using these manipulations,
so these manipulations define the type of attack. After attacking the database Mallory yields a new database.

%

Alice has to make her watermark robust against these attacks so that they cannot affect the embedded watermark. These attacks can be broadly classified as:

\begin{itemize}
  \item \textbf{$A_{1}$}(Insertion attacks): Mallory may insert $\alpha$ new records in the watermarked database.
  \item \textbf{$A_{2}$}(Deletion attacks): Mallory may delete $\alpha$ records from the watermarked database.
  \item \textbf{$A_{3}$}(Alteration attacks): Mallory may alter $\alpha$ records from the watermarked database.
  \item \textbf{$A_{4}$}(Multifaceted attacks): Mallory may delete $\alpha$ records, alter $\beta$ records, and
  insert $\gamma$ new records to form a new database.
\item \textbf{$A_{5}$}(Additive (Re-watermarking) attacks): In such attacks, Mallory inserts his own watermark in the already marked database.

\end{itemize}

Similarly, for fragile watermarking technique, the major requirement for the watermark is to be vulnerable to even minor modifications of the dataset while also fulfilling the additional requirements of characterization and localization of attacks.

After the pioneer work by Agrawal et al. \cite{agrawal2002watermarking} the problem of
  relational databases watermarking has been addressed in several techniques, with each technique having its own benefits, drawbacks,
issues and limitations. Some of these techniques are reversible (for instance, \cite{jawad2013genetic}, \cite{linovel}, \cite{gupta2008reversible}) while other are irreversible (for instance, \cite{kharadespeech}, \cite{gao2013new}, \cite{franco2015robust}, \cite{kamran2013robust}) depending upon the nature of their intended applications. Similarly, some techniques are meant for ownership protection  (for instance, \cite{jin2006watermarking}, \cite{khanduja2015watermarking}, and \cite{franco2014robust}) while others are designed for data integrity and verification (for instance, \cite{camara2014block} and \cite{khan2013fragile}). For database watermarking, usually bit strings \cite{sion2004rights},  images \cite{al2010copyright}, speech signals \cite{wang2008speech}, character strings \cite{yong2006novel} are used as a watermark.

In \cite{halder2010watermarking} Halder et al. presented a survey for classification and comparison of some
of these watermarking techniques. We believe that a detailed survey of relational database watermarking techniques is an important need in order to facilitate researchers to explore this area of research by providing them access to state-of-the-art research. This paper presents a comprehensive survey of relational database watermarking and
fingerprinting techniques -- developed to date -- with the objective of developing an understanding about: (1) how and where the watermark is inserted; (2) their resilience against malicious attacks; (3) the shortcomings of
the existing work. Moreover, the paper will also act as a reference to provide pointers to latest research in this field. As a result, it will enable researchers to design an effective watermarking technique for their application domain.

We classify the existing work done into three broad
 categories based on how and where the watermark is inserted. We also present a comparison of major watermarking techniques -- belonging to the same classification hierarchy in our paper -- so that a reader can learn about their evolutionary history. For this purpose, we short-listed only those papers for comparison which were published in reputable computer science Conferences and Journals\footnote{The list of computer science Conferences and Journals ranking can be found at http://academic.research.microsoft.com/?SearchDomain=2\&entitytype=2 (link last accessed on April 04, 2016).} and for brevity, we also did not compare all of them.

For collecting the research articles related to relational database watermarking, we have used the concept of citation graph by making two way transitions: (1) the first one is to select a pioneer paper and look at relevant references in its list; and (2) use Google Scholar to identify all articles (published after this article) that cite it. As a result, we have short listed more than 100 papers. The literature search was exhaustive and every effort is made to include papers published in prestigious Journals and Conferences.

\begin{figure*}[htp]\scriptsize
\centering \includegraphics[angle=0,
width=2.0\columnwidth]{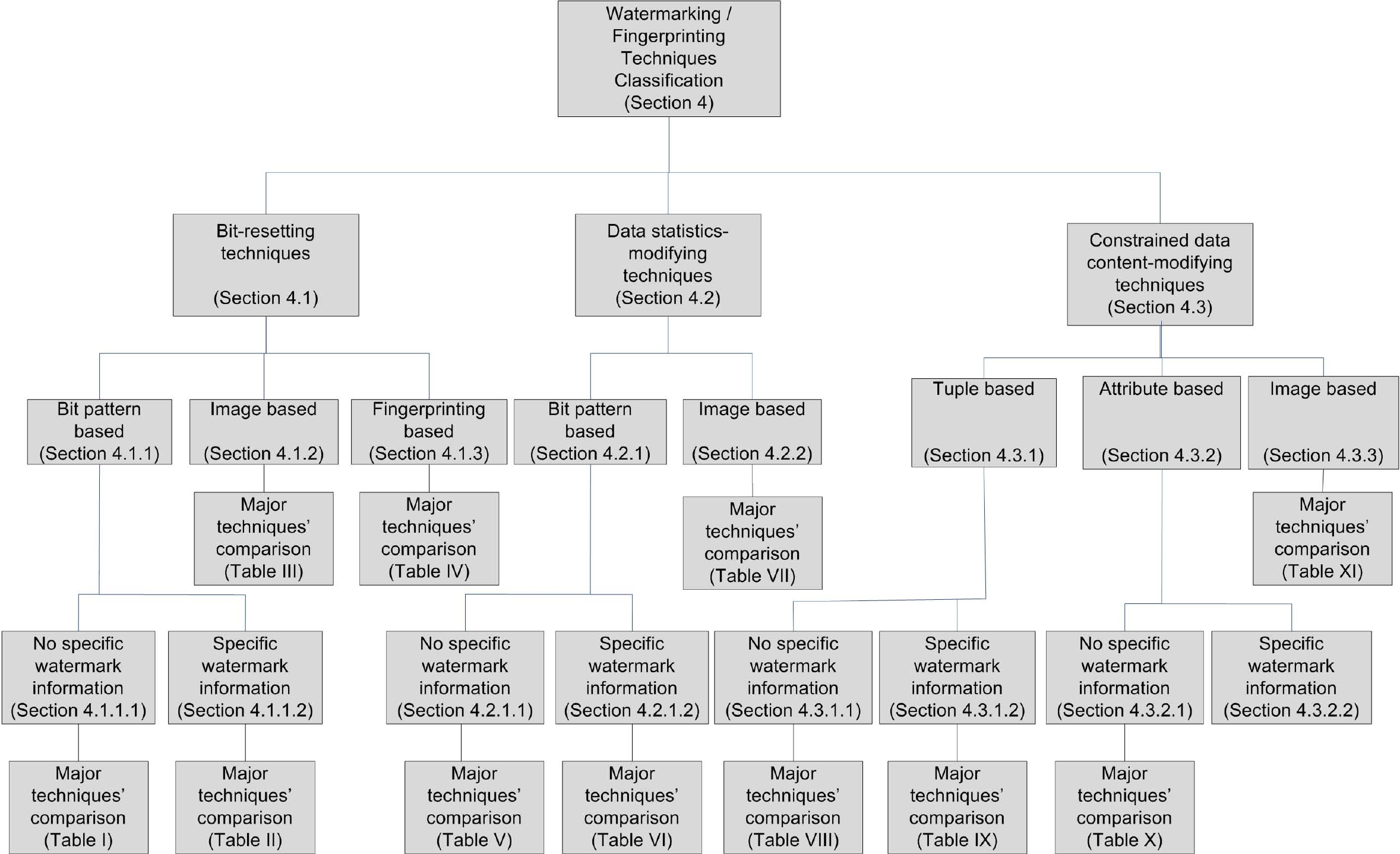} \caption{Diagram for depicting hierarchical structure of review and classification of watermarking and fingerprinting techniques.}\label{fig:PaperLayout}
\end{figure*}

We have organized the rest of the paper as follows. We present the brief description of the related work in Section
\ref{sec:RelatedWork}. A generic framework of relational database watermarking technique is given in Section \ref{sec:GenericFramework} and a comprehensive review and classification of watermarking and fingerprinting techniques in
Section \ref{sec:TechniquesClassification} (for ease of reference figure \ref{fig:PaperLayout} shows the hierarchical structure of review and classification of watermarking and fingerprinting techniques). Then we give a comparison of different classes of watermarking schemes and point out the future directions in Section \ref{sec:ComparisonofTechniques} and conclude in Section \ref{sec:Conclusion}.

\section{Motivations and Related Work}
\label{sec:RelatedWork}
The major motivation behind this work is to develop a better understanding of the relational data watermarking techniques for real world problems. Moreover, we are also interested in the vulnerability of these techniques against malicious attacks. We also analyze the process of defining data usability constraints
and their relevance to relational data.

 In \cite{sion2004attacking}, Sion et al. define some attacking scenarios available for the attacker and suggest design choices to increase watermark safety but they did not investigate the attacker model in detail and also their objective was not to present a comparison of watermarking techniques.  In \cite{halder2010watermarking}, Halder et al. presented a classification and comparison of some relational database watermarking techniques. In \cite{lafaye2007analysis}, authors presented the idea of two models regarding leakage of secret watermarking contents. These models are based on different combination of keys and databases. In \cite{rathva1watermarking}, \cite{mehta2014watermarking}, and \cite{iftikhar2015survey} the survey of few selected database watermarking techniques has been presented.

But the focus of these efforts are also not towards comparing different watermarking techniques under different attack scenarios. On the other hand, the focus of our work is to present the first (comprehensive) survey by critically reviewing almost all relational database watermarking techniques -- presented so far -- by: (1) highlighting issues related to watermark embedding; (2) investigating their robustness; (3) exploring their applicability for real world problems; and (4) suggesting the future directions of research.
\section{A Generic Framework for Relational Database Watermarking/ Fingerprinting Techniques}
\label{sec:GenericFramework}
\begin{figure*}\scriptsize
\centering \includegraphics[angle=0,
width=1.5\columnwidth]{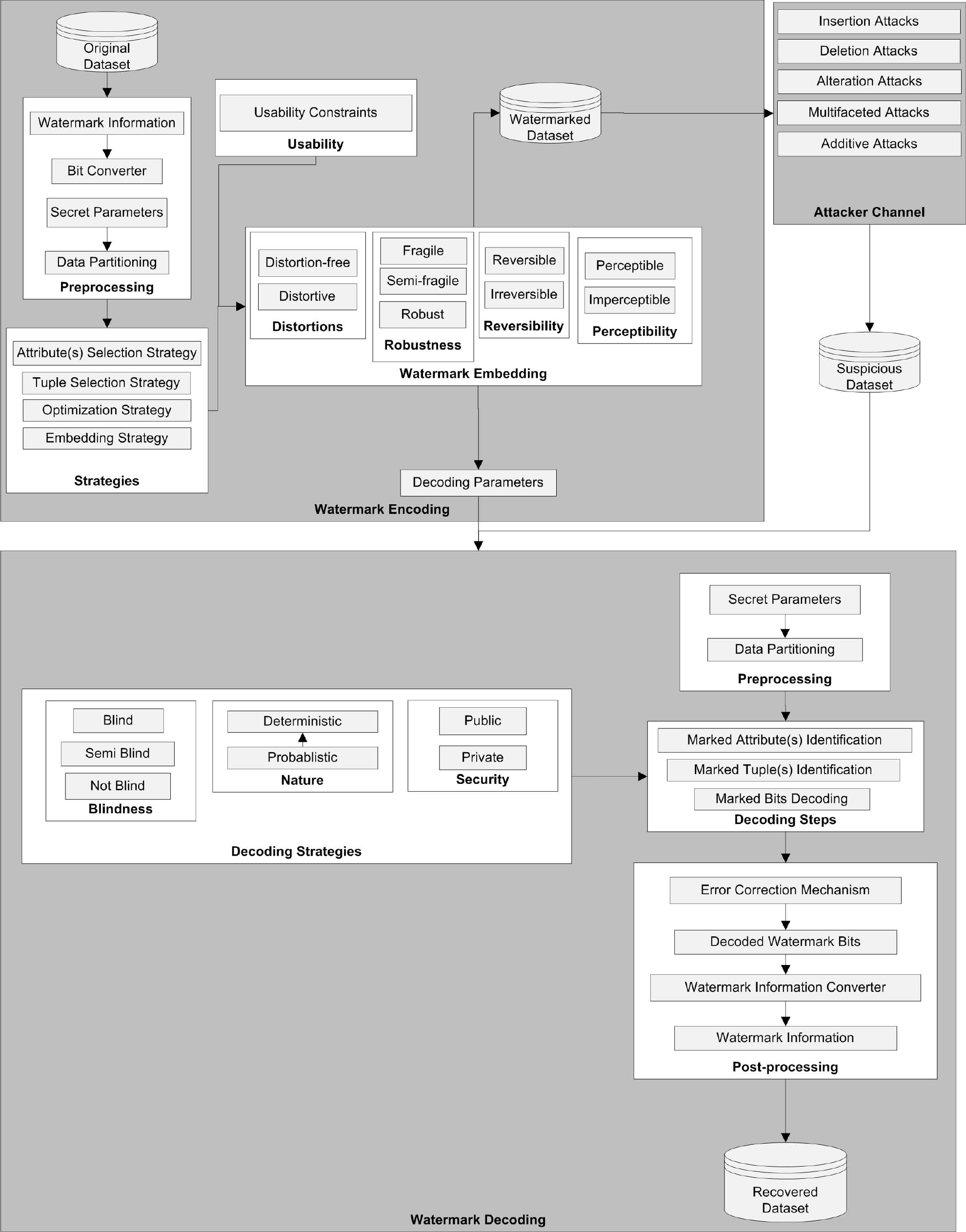} \caption{Diagram for depicting different components of a watermarking technique.}\label{fig:DBWMFramework}
\end{figure*}

The watermarking or fingerprinting technique needs to meet certain common requirements, however, they do still differ in many other aspects. This section entails a brief description of these common requirements and also highlights different decision choices  among the watermarking (and fingerprinting) techniques. The advantage of this description is two folds: (1) it provides a generic unified reference framework for describing and comparing different watermarking and fingerprinting techniques; (2) this framework can also be used as a guideline for developing meta-techniques for watermarking relational data.
It consists of three modules and a several submodules that can be used to design and implement a watermarking technique. The top level modules include: (1) watermark encoding; (2) attacker channel; and (3) watermark decoding. The framework in figure \ref{fig:DBWMFramework} depicts the characteristics of the different modules and their inter-module and intra-module collaborations. The functionality and characteristic of each module and its subsequent components is given in the next subsection.

\subsection{Watermark Encoding}
\label{sec:WatermarkEncoding}
The main purpose of this module is embedding ownership information in the dataset. This module may include: (1) preprocessing steps like data partitioning, defining secret parameters etc.;  (2) strategies to select data for watermarking while using an optimization strategy to ensure best watermark encoding; (3) defining usability constraints to ensure quality of data during watermark embedding; (4) embedding watermark subject to usability constraints; (5) computing a number of parameters for use in watermark decoding stage, and finally (6) generating the watermarked dataset for delivering it to the intended recipients. The brief description of each submodule is given in the following.

\textbf{Preprocessing.} The optional preprocessing steps may involve conversion of watermark information (images, owner name etc.) to bit strings. The data is also usually divided into logical groups using a particular given criterion. Data partitioning is performed using secret parameters to ensure that only the data owner knows the information about the data partitioning.

\textbf{Strategies.} Strategies for selecting attribute(s), tuple(s), watermark optimization, and watermark embedding are different components of this sub-module. It is not mandatory that a watermarking technique will utilize all of the above-mentioned strategies.

\textbf{Usability.} Usability constraints are defined by the data owner to identify bandwidth for watermarking. The sole purpose of this sub-module is to ensure that the watermarked data remains usable.

\textbf{Watermark embedding.} This sub-module embeds the watermark in accordance with the encoding strategies and data usability. The watermark embedding has to cater a number of challenges by answering questions like whether: (1) an inserted watermark brings distortions in the original data; (2) an inserted watermark is fragile (watermark is corrupted after even minor malicious attacks), semi-fragile(watermark may tolerate minor malicious attacks, but cannot tolerate major attacks), or  robust against attacks (can tolerate malicious attacks to a certain level); (3) the inserted watermark is reversible or irreversible (that is original data may or not be recovered after watermark decoding); (4) an inserted watermark is perceptible or imperceptible (but it is desired that relational database watermarking should be imperceptible).

\textbf{Decoding parameters}

Some of the parameters of watermark embedding are usually required for watermark decoding, that are calculated and saved by this submodule.

\textbf{Watermarked dataset}

The main output of watermark encoding module is a watermarked dataset that subsequently delivered to intended recipients.

\subsection{Attacker Channel}
\label{sec:AttackerChannel}
An attacker -- Mallory -- can launch a number of malicious attacks on the watermarked data. The attacker channel module refers to all such possible attacks. The attacks can modify some contents of the data with an aim to disturb the embedded watermark. The description of this module has already been given in Section \ref{sec:Introduction}.
\subsection{Watermark Decoding}
\label{sec:WatermarkDecoding}
This module is used to decode the embedded watermark from the suspicious data. Usually, it involves: (1) some preprocessing steps for dataset partitioning (if it was performed during watermark embedding); (2) decoding strategies for recovering watermark; (3) decoding steps for marked attribute(s), marked tuple(s), and marked bit identification; (4) error correction and recovery steps; and (5) recovery of original data (for reversible watermarking techniques only). The brief description of each component of this module is as follows.

\textbf{Preprocessing.} If dataset partitioning was performed during watermark encoding, the same steps are repeated to generate data partitions.

\textbf{Decoding Strategies.} The watermark decoding strategies include: (1) blindness of watermark decoding algorithm -- a blind decoding algorithm does not require any part of original data or embedded watermark; but semi blind decoding usually requires the embedded watermark during decoding; (2) the watermark decoding may be probabilistic or deterministic in nature; and (3) watermark detection may be public or private depending on the security requirements.

\textbf{Decoding steps.} The marked attribute(s) and tuple(s) are identified and watermark bits are decoded by this submodule.

\textbf{Post processing.} The main responsibility of this submodule is to employ error correction mechanism (if used) on the decoded watermark bits, and convert these bits into the watermark information that was embedded as the watermark (ownership information).

\textbf{Recovered dataset.} After watermark decoding the original data may be recovered by reversible watermarking techniques only.

\section{Watermarking/Fingerprinting Techniques Classification}
\label{sec:TechniquesClassification}
\begin{figure*}[htp]\scriptsize
\centering \includegraphics[angle=0,
width=0.8\columnwidth]{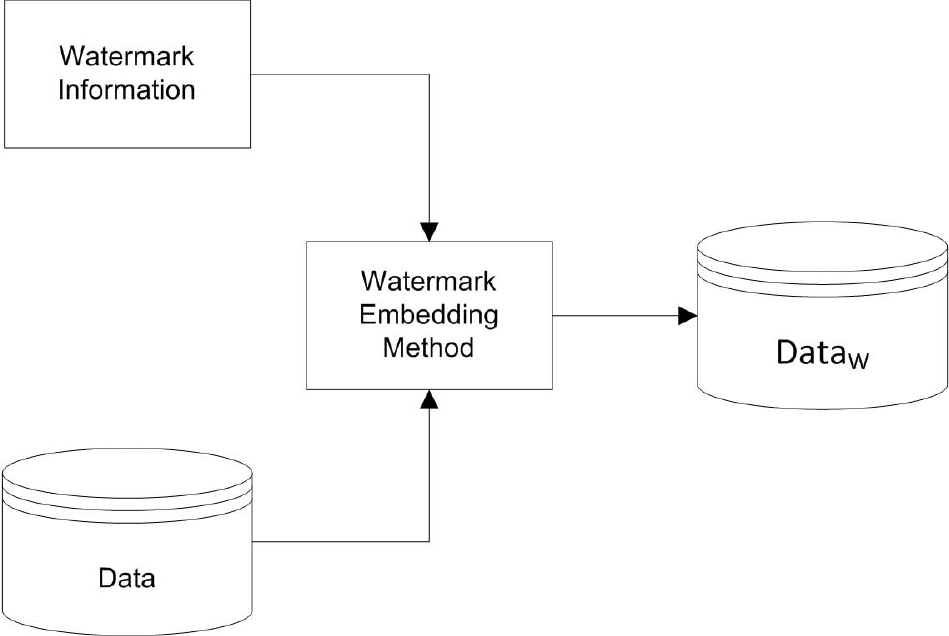} \caption{The building block for classification of watermarking  techniques.}\label{fig:WMTechClassification}
\end{figure*}

Figure \ref{fig:WMTechClassification} is used to provide a basis for classification of watermarking and fingerprinting techniques. In this figure the blocks containing label "Watermark Information" and "Watermark Embedding Method" are used to define different classes and their hierarchies. In this paper, we classify the watermarking/fingerprinting techniques into three broad categories based upon these two blocks. They are:

\begin{itemize}
  \item Bit-resetting techniques(BRT). In these techniques, selected bits (mostly LSBs) are reset by following a systematic
  process.
  \item Data statistics-modifying techniques(DSMT): Data statistics such as mean, variance, distribution are used to embed watermark in these techniques.
 \item Constrained data content-modifying techniques(CDCMT). These techniques are based on modifying the contents of the data, for example, techniques based on the ordering of the tuples(such as in \cite{bhattacharya2009distortion}), insertion of extra spaces in attribute values  (such as in \cite{al2008robust}) such that watermarked data still remains useful. We also place the zero-watermarking techniques (like \cite{Hamadou2011fragile}) under this category.

\end{itemize}
Now we give a review of techniques that belong to these top level techniques.

\subsection{Bit-resetting techniques}
We further classify such techniques into three categories:
\begin{itemize}
  \item Bit pattern based bit-resetting techniques(BPBBRT): In such techniques a bit pattern is used as a watermark and embedded in the selected
  bits of the data.
  \item Image based  bit-resetting techniques(IBBRT): Image is used as a watermark, for embedding in the selected
  bits of the data, in such techniques.
 \item Bit-resetting fingerprinting techniques(BRFT): These techniques employ fingerprinting phenomenon for marking selected bits of the
 relational databases.

\end{itemize}

\subsubsection{Bit pattern based bit-resetting techniques}
%

\paragraph{BPBBRT with no specific watermark information}

In \cite{agrawal2002watermarking} Agrawal and Kiernan proposed a pioneering work of watermarking relational databases. A parameter $\gamma$ is used to compute the fraction of tuples to be watermarked. In watermark embedding phase, a tuple
is selected for watermarking if the hash value of its primary key modulo $\gamma$ is equal to zero. In the next step, an attribute and the bit position to be watermarked is selected. A watermark bit is embedded in a tuple by computing a hash value after applying a hash function on primary key and a secret key. If this hash value is even, $j^{th}$ LSB  of the attribute values is set to 0;
otherwise, it is set to 1. The watermark detection algorithm starts by identifying marked tuples, the marked attribute and marked bit. This bit is matched with the embedded bit and a threshold for matching bit count is computed. If the match count is greater than or
equal to this threshold, watermark detection is said to be successful. To extend this work, Agrawal et al. in \cite{agrawal2003system}, \cite{agrawal2003watermarking} used a pseudo-random sequence generator to select tuples for watermarking with a user specified \emph{gap parameter}. The authors did not use any mechanism for data usability after watermark embedding and used one bit watermark. As a result, such techniques are vulnerable
to simple attacks. For instance, shifting of only one bit results in deletion of watermark bits without a significant loss of data usability. As a result, such techniques are not well-suited for datasets which can tolerate changes in the LSB positions. A number of recent techniques like \cite{xiao2007second}, \cite{manjula2010new}, \cite{Hamadou2011Weight}, \cite{khanduja2012identification}, \cite{zhang2011relational}, \cite{rao2012subset}, \cite{iqbal2012self}, \cite{yanmin2012digital} extend the work of \cite{agrawal2002watermarking} and embed multi-bit watermark in selected LSBs; while, \cite{tiwari2013novel} embeds watermark in third most significant bit (MSB) of the numeric attribute. Similarly, in \cite{hu2005garwm} the candidate bit for watermarking is computed.


The watermarking techniques in \cite{sion2004proving} and \cite{sion2005rights} provided mechanism for providing ownership protection
for categorical data. The tuples "fit" for watermarking are selected by using a message authentication code (MAC) which also takes a secret key
$k_{1}$ as an input parameter. In the next step, a secret value for a categorical attribute in "fit" tuples is generated and its
LSB is set to a value based on securely computed position. This position is computed by another secret key $k_{2}$. In decoding phase,
the "fit" tuples from the watermarked data are again computed using the same procedure as in the watermark encoding phase. Watermark bits
are then decoded by setting $wm\_data[msb(H(T_{j}(K), k_{2}),b(\frac{N}{e}))]=t\&1$; where $wm_data$ is watermarked table, $T_{j}$ denotes the $j^{th}$ tuple, $K$ is
the primary key, $N$ is the mapping of categorical data into bit strings, $e$ determines the tuples considered for watermarking, and $t$
represents a bit value. An error correction code (e.g., majority voting) is also applied to eliminate the watermark decoding errors. Though these are first major contributions towards watermarking of categorical attributes but these techniques are not suited for sensitive categorical data (e.g. medical data) where changing the value of a categorical attribute may lead to incorrect information regarding an important entity. Moreover, if data is updated, the watermark may be disturbed specially if a query alters a large number of tuples. For numeric data, Cui et al. in \cite{cui2006robust} modified the technique of \cite{sion2005rights}.

In \cite{huang2004new}, a unique $id$ is calculated for each tuple of the table using a
hash function. The next step is to partition the data into $p$ partitions with each partition having the same number of records and then mark them subject to usability constraints. The usability constraints include statistical measurement constraints, semantic constraints and
structural constraints. The authors use mean and variance of the data as statistical usability constraints while the semantic and structural
constraints are specified by SQL statements. If these constraints are violated for any selected tuple, that tuple is excluded during
watermark detection phase to improve watermark decoding accuracy. In watermark decoding stage, the tuples's $id$s are again computed using
the same hash function and data is partitioned. The bits are detected from each partition probabilistically and the majority voting is used as error correction mechanism. The technique suffers from synchronization errors due to insertion and deletion of new tuples because $id$s
of tuples forming the partition boundaries will no longer remain the same as during the watermark encoding stage.



In \cite{bertino2005privacy}, Bertino et al. proposed a hierarchical watermarking for outsourced medical data. They use the similar approach for watermarking as in \cite{sion2005rights} after binning the medical data and embed the watermark in different hierarchies.
The hierarchies are defined with the help of user roles. Generalization (trees)nodes are used to depict these hierarchies.


In a fragile watermarking technique proposed in \cite{guo2006fragile}, watermark embedding is done in such a way that it forms a grid -- which aids
in identifying modifications made to the watermarked database relation. During watermark decoding stage, tuples are again grouped
and two verification vectors are formed for each group to detect the two embedded watermarks. These two vectors are also used to
identify any modifications made to the watermarked data.

In \cite{Cui2008Public} Cui et al. proposed a watermarking technique using combination of private key and public key. A trusted
third party $IPR$ is informed about the public key and the data owner is the only one who knows the secret private key. In watermark
insertion phase, a $mixed \phantom{1} code$ is formed using the public key and the private key. For watermark embedding, the appropriate bit
position is determined using the hash function and the LSB of the attribute is marked. During watermark decoding, the $mixed \phantom{1} code$ is computed again and the majority voting scheme is applied to correctly locate the marked bits. Finally, a map transformation is used to decode
the embedded watermark.

Wang et al. \cite{wang2008watermarking} applied $v$ different watermarking schemes on a database relation $R$ as a preprocessing step and
then used the best one for watermark embedding. A scheme is selected on the basis of its impact on the data -- the lower impact is preferred in the objective function.

Meng et al. \cite{Meng2008approach} used genetic algorithm (GA) for generation of watermark. The chromosome is a bit string of zeros and ones.
The best chromosome is used as a watermark. The watermark encoding and decoding algorithm is same as proposed in \cite{cui2006robust}.

A shadowed watermark based technique was proposed in \cite{xian2009leakage} by Xian et al. The data is
watermarked with a secret watermark key and a secret shadowed key which is different for every data user. A trusted watermark server (TWS)
watermarks the data and assigns the shadowed key to the data user. When illegal data leakage is detected, the shadowed key of the data user
is used to identify the actual source of data leakage; as a result, the innocent user is not falsely accused of data leakage. Although this techniques tends to protect an innocent user but it also suffers from the same shortcomings as suffered by the technique presented in \cite{agrawal2003watermarking}.

Table \ref{tab:BPBBRTsMeaningLess} shows the summary and a comparison of well known BPBBRTs with no specific watermark information. This table compares different techniques based on certain parameters depicted in Figure \ref{fig:DBWMFramework}. Note that the following nomenclature has been used in tables  \ref{tab:BPBBRTsMeaningLess} through \ref{tab:IBCDCMT}.

\textbf{Column 1:} Sch. provides a reference to a watermarking technique.

\textbf{Column 2:} MAT refers to marked attribute data type. This column can have values:
All (all data types), N (numeric data), C (categorical data), and AN (alpha-numeric data).

\textbf{Column 3:} ASM is acronym for attribute selection method and refers to the procedure of selecting attribute(s) for watermarking. This column can have values: Arbitrary (attribute(s) for watermarking is/are arbitrarily chosen by data owner), Weighted (based on weights of attribute(s) weight), PRSG based (pseudo-random sequence generator (PRSG) based), All (all attributes are selected for watermarking), SHF based (secure hash function (SHF) based), Property (based on some particular properties of attribute), information gain (\emph{IG}) based, mutual information (\emph{MI)} based, and Index based (database Index based).

\textbf{Column 4:} TSM stands for tuple selection method and refers to the procedure of selecting tuple(s) for watermarking. This column can have values: Arbirary (tuple(s) for watermarking is/are arbitrarily chosen), PRSG based (pseudo-random sequence generator (PRSG) based), All (all tuples are selected for watermarking), SHF based (secure hash function (SHF) based), and Index based (Database Index based).

\textbf{Column 5(for robust techniques):} GL is granularity level and refers to the basic unit where the watermark is embedded. This column can have values: Bit (bit level), Value (attribute value),  Tuple (tuple level), Table (database table level), and
Index (database index level).

\textbf{Column 5(for fragile techniques):} Loc. refers to localization levels (if any) to which a fragile technique is able to locate the tempered data. This column can have values: Group (referring to a data block or partition level), Tuple(record level),  , and - (referring to the fact that the technique is not able to localize the data tempering attacks).

\textbf{Column 6:} DMB shows whether a decoding method is blind or not. This column can have values:
Blind, Semi-blind, and Not blind (original data is required during watermark decoding).

\textbf{Column 7:} ECM is abbreviation of error correction mechanism and refers to error correction mechanism adapted by the watermark decoding scheme. This column can have values: - (no error detection mechanism is used by the decoding scheme), Voting (the majority voting scheme is used for error detection),
CRC	(the cyclic redundancy check is used as an error detection mechanism), and EPC (the even parity check is used as an error detection mechanism).

\textbf{Column 8:} Dep. refers to dependencies (if any) of the watermark decoding scheme for successful watermark decoding. This column can have values:
\begin{itemize}
  \item PK: The successful watermark decoding depends on primary key,
  \item Order: The  successful watermark decoding depends on the order of tuples in the watermarked database, if the order of tuples is changed then the watermark decoding accuracy may decrease,
 \item Value: The  successful watermark decoding depends on the watermarked attribute(s) value, if the value of the attribute(s) is changed then the watermark decoding accuracy may decrease,
 \item Markers: The  successful watermark decoding depends on special marker tuples which are used to identify the boundaries of data partitions in the watermarked database, if the position of these marker tuples is changed then the watermark decoding accuracy may decrease,
 \item Unique Column: The successful watermark decoding depends on unique identifying column(s), if the value of the unique identifying column(s) is changed then the watermark decoding accuracy may decrease.
 \item Information gain (IG): The  successful watermark decoding depends on information gain of the attributes, if the value of the information gain of attributes is changed then the watermark decoding accuracy may decrease.
\end{itemize}

\textbf{Column 9:} Dist. is used for to show distortions introduced by watermark embedding scheme. This column can have values: Yes (the watermark embedding results in distortions in original data), and No (the watermark embedding does not introduce distortions in original data).

\textbf{Column 10:} Opti. refers to any optimization strategy used in the proposed technique. This column can have values: Yes, and No.


\textbf{Column 11 (for robust techniques):} Rev. refers to the fact that whether the proposed technique is reversible or not. This column can have values: Yes(reversible), and No(irreversible).

\textbf{Column 11 (for fragile techniques):} Char. refers to the fact that whether the proposed fragile technique is able to identify (characterize) the type of attack (tuple insertion, deletion, and alteration attack). This column can have values: Yes(the technique is able to characterize the attack), and No(the technique is not able to characterize the attack).


\begin{table*}[htbp]\tiny
  \centering
   \caption{Summary and comparison of well known BPBBRTs with no specific watermark information.}
     \label{tab:BPBBRTsMeaningLess}%
    \begin{tabular}{|l|l|l|l|l|l|l|l|l|l|l|}
       \hline
  \multicolumn{11}{|c|}{\textbf{Robust Techniques}}\\
    \hline
   \textbf{Sch.} & \textbf{MAT} &\textbf{ASM} & \textbf{TSM} & \textbf{GL} & \textbf{DMB} &\textbf{ECM} & \textbf{Dep.} & \textbf{Dist.}& \textbf{Opti.}&  \textbf{Rev.} \\
   \hline
    \cite{agrawal2002watermarking} & N & Arbitrary & SHF & Bit &Blind & -& PK, & Yes& No & No \\
     &  &  & based &  &  &  &  Value& &  &  \\
      \hline
   \cite{agrawal2003watermarking} & N     & Arbitrary & {PRSG} & Bit   & Blind & -     & PK, & Yes & No & No \\
          &       &  & based &       &       &       &       Value& &              &       \\
           \hline
   \cite{sion2005rights} & C     & Arbitrary & SHF & Bit   & Blind & Voting & PK, & Yes& No & No \\
          &       &  & based &       &       &       &       Value&  &       &\\
           \hline
   \cite{xiao2007second} & N     & Arbitrary & SHF & Bit   & Blind & {PRSG} & PK, & Yes& No & No \\
          &       &       & based &       &       & based &       Value&&       &       \\
           \hline
   \cite{wang2008watermarking} & N     & Arbitrary & SHF & Bit   & Blind & Voting & PK, & Yes& No & No \\
          &       &       & based &       &       &       &       Value&  &       &\\
    \hline
  \multicolumn{11}{|c|}{\textbf{Fragile Techniques}}\\
   \hline
   \textbf{Sch.} & \textbf{MAT} &\textbf{ASM} & \textbf{TSM} & \textbf{Loc.} & \textbf{DMB} &\textbf{ECM} & \textbf{Dep.} & \textbf{Dist.}& \textbf{Opti.}&  \textbf{Char.} \\
   \hline
   \cite{guo2006fragile} & N     & SHF & SHF & Tuple   & Blind & -     & PK, & Yes & No & Yes\\
          &       & based & based &       &       &       &       Value&  &       &\\
    \hline
    \end{tabular}%
\end{table*}%

\paragraph{BPBBRT with specific watermark information}

In \cite{zhang2006relational}, Zhang et al. used the database content characteristics for watermarking relational databases. In this technique, a tuple is selected for watermarking based on a random number between 0 and 1.
A function $InterString(Ab_{i}, K, S)$ is used to extract a substring from $Ab_{i}$ (binary equivalent of the selected numeric attribute for
watermarking), where $K$ is a secret key, $S$ is the length of substring. The mark bit is embedded at the end of the
binary value of attribute $A_{2}$. In watermark detection phase, $InterString$ is again computed and the mark bit is matched with the last bit of
$A_{2}$; if the mark bit is correctly detected, a match count is updated. If this match count is greater than a certain threshold, the watermark is successfully detected. The decoding accuracy of this scheme suffers if an attacker by launching alteration attacks alters the attribute value and as a consequence its binary representation.

Qin et al. \cite{qin2006watermark} used the idea of chaotic random numbers for embedding watermarks in relational database to overcome the shortcomings of \cite{agrawal2003watermarking}. Authors generate
these numbers by using logistic equation and process only the integral part of these numbers. The logistic chaos equation (LCE) used is $X_{n+1} = \mu x_{n}(1-x_{n})$; where, $\mu \epsilon [1,4]$ and $n$ is the number of tuples in the database.

In \cite{guo2006improved}, Guo et al. converted a meaningful watermark to a bit string.
The authors then locate the candidate attributes and candidate bits. The database is partitioned into variable but similar sized groups. The $i^{th}$ bit of
the watermark is embedded into tuples of $i^{th}$ group.


The authors have also proposed a recovery algorithm, if watermark is located in the data. For this purpose, the detected bits for the subsets  $subset\_0$ (zero bits) and  $subset\_1$ (one bits) are counted
and the bit having larger $total\_count$ is supposed to be actual embedded bit.

In \cite{gupta2009database}, a technique addressing additive attacks on the watermarked relational databases has been proposed. In the watermark insertion phase of this technique, a real valued attribute is converted into its binary equivalent and a
bit is extracted from the integral part of a real valued attribute ($A$). This bit is then embedded as the watermark into fractional part of the
same attribute if $(len(int(A))) > (pos_{1} + \epsilon)$ and $len(frac(A)) < cap$, where, $pos_{1}$ is the LSB of the integer part at position 1,
$\epsilon$ is a parameter used to control the amount of distortion introduced by bit embedding and $cap$ is the number of bits allocated for
fraction part of the number. In the watermark decoding stage, the value of the fractional part is stored by locating the watermark bit in the fractional
part and replacing it by the $oldBit$ which was marked in the watermark embedding phase. The secondary watermark attacks are handled by computing
the watermarks of each party (claiming the ownership of the data) and a party $p_{i}$ is considered as the legitimate owner of a data
$x_{i}$ if and only if their watermark is detected in $x_{i}$. This technique is able to provide a good solution to an important problem of additive attacks but is not very resilient to other attacks.



Other such techniques are presented in \cite{cui2007weighted}, \cite{ali2011watermarking}, \cite{wang2008speech},  and \cite{farfoura2012blind} but for brevity, we omit their details because they apply small modifications to the techniques discussed above.

Table \ref{tab:BPBBRTsMeaningfull} shows the summary and comparison of well known BPBBRTs with specific watermark information based on different parameters depicted in Figure \ref{fig:DBWMFramework}.

\begin{table*}[htbp]\tiny
  \centering
   \caption{Summary and comparison of well known BPBBRTs with specific watermark information.}
     \label{tab:BPBBRTsMeaningfull}%
    \begin{tabular}{|l|l|l|l|l|l|l|l|l|l|l|}
       \hline
  \multicolumn{11}{|c|}{\textbf{Robust Techniques}}\\
    \hline
   \textbf{Sch.} & \textbf{MAT} &\textbf{ASM} & \textbf{TSM} & \textbf{GL} & \textbf{DMB} &\textbf{ECM} & \textbf{Dep.} & \textbf{Dist.}& \textbf{Opti.}&  \textbf{Rev.} \\
   \hline
    \cite{qin2006watermark} & N     & All   & Arbitrary & Bit   & Blind & -     & PK, Value & Yes& No &  No \\
     \hline
   \cite{guo2006improved} & N     & Arbitrary & SHF   & Bit   & Blind & Voting & PK, Value & Yes& No &  No \\
          &       &       & based &       &       &       &       & &              &         \\
           \hline
    \cite{gupta2009database} & N     & Arbitrary & SHF   & Bit   & Not Blind & -     & PK, Value & Yes& No &  No \\
     \hline
    \end{tabular}%
\end{table*}%

\subsubsection{Image based bit-resetting techniques}
%

Yong et al. in \cite{zhang2005Amethod} presented a image based watermarking technique to verify
ownership on relational databases. First a subset of tuples is selected for watermarking using a
secure hash function and then ordered $RGB$ values of the image are embedded as a watermark in the database relation.
For watermark detection, the same hash function is applied to identify the marked tuples and embedded pixel values are decoded. A similar approach was recently proposed in \cite{rao2012relational}.

Zhou et al. \cite{zhou2007additive} used BMP image as a watermark in the database.  The BMP image is first split into two parts: the header
part and the main information of the image, i.e., data array. Bose-Chaudhuri-Hocquenhem (BCH) \cite{shankar1979bch} code is
then generated as a watermark from the data array of the image. Authors show the resilience of proposed technique again additive-attacks. They suggest the idea of a trusted third party (TTP) to investigate about the actual owner of the database.


In \cite{wang2008atbam}, Wang et al. proposed a watermarking technique for relational data by applying Arnold transform \cite{arnol1968ergodic} on an image with data partitioning based watermark embedding and decoding algorithms. This technique is vulnerable to tuple insertion, alteration and deletion attacks. Furthermore, a small bit flipping attack effects the watermark decoding accuracy as the modulo operation yields different results.


Hu et al. also presented an image based watermarking algorithm for relational databases in \cite{hu2009image}. Almost similar
work is presented by Sun et al. in \cite{sun2008multiple}. They transform the image into a bit flow which is used as a watermark. In \cite{sardroudi2010new}, authors use image elements to embed an image-based watermark in the database relation.
 Like \cite{wang2008atbam}, such techniques technique also face the same challenges of robustness.

The technique in \cite{zhang2011relational} also use image watermark for textual attributes and numerical attributes. In \cite{chang2012reversible}, authors use SVR along with FP-tree for reversible database watermarking. Another such technique was proposed in \cite{khanduja2015watermarking} where authors use image (other files such as audio can also be used for watermarking) to create watermark bits and then embed the same using a partitioned based approach.



Table \ref{tab:IBBRTs} shows the summary of well known IBBRTs based on certain parameters shown in Figure \ref{fig:DBWMFramework}.

\begin{table*}[htbp]\tiny
  \centering
   \caption{Summary of well known IBBRTs.}
     \label{tab:IBBRTs}%
    \begin{tabular}{|l|l|l|l|l|l|l|l|l|l|l|}
       \hline
  \multicolumn{11}{|c|}{\textbf{Robust Techniques}}\\
    \hline
   \textbf{Sch.} & \textbf{MAT} &\textbf{ASM} & \textbf{TSM} & \textbf{GL} & \textbf{DMB} &\textbf{ECM} & \textbf{Dep.} & \textbf{Dist.}& \textbf{Opti.}&  \textbf{Rev.} \\
   \hline
   \cite{zhou2007additive} & N     & Arbitrary & SHF   & Bit   & Semi Blind & Voting & PK, Value & Yes& No &  No \\
          &       &       & based &       &       &       &       &  &&\\
    \hline
    \cite{khanduja2015watermarking} & N     & Arbitrary & SHF   & Bit   & Blind & Voting & Value & Yes& Yes &  No \\
          &       &       & based &       &       &       &       &  &&\\
    \hline
    \end{tabular}%
\end{table*}%

\subsubsection{Bit-resetting fingerprinting techniques}
%

Fingerprinting techniques in \cite{li2003constructing} embed a meaningful fingerprint of the data to protect the interests of buyer.
The authors used the MAC on a secret key and a user(buyer) identifier to compute the fingerprint of length $L$ with $L > logN$ and $N$ being the number of buyers. Database relations generally
have a primary key attribute which may be used (along with other required parameters for hash function) for calculating
the fingerprints. The authors also propose three different approaches to fingerprint relations that do not have any primary key.
In their first approach(S-Scheme), the authors have considered the candidates(for watermarking) bits of a single numeric attribute as the
virtual primary key. This approach suffers from two main problems: (1) The candidate bits may not be unique for each tuple
(duplicate problem); and (2) if the virtual primary key is dropped by the attacker, the fingerprint information is lost
(deletion problem). In their second approach(E-Scheme), Li et al. investigate the value of each numeric attribute in a tuple. Next, a virtual primary key is constructed from each attribute. This approach solves the deletion problem but the duplicate problem
still persists in this approach. In the third approach(M-Scheme), bit positions are dynamically selected to construct a virtual primary key. In this approach, different attributes are selected for different
tuples and hence duplication problem is resolved. Moreover, the fingerprint is not embedded in only one attribute; therefore, dropping selected attributes does not result in the fingerprint deletion problem.

In \cite{li2005fingerprinting}, Li et al. have selected tuples for fingerprinting using a hash function. The mark embedding and detection scheme is same as proposed in \cite{zhou2007additive}.

Liu et al. \cite{liu2005block} used blocks of bits that were available for marking to embed for fingerprint.


In \cite{constantin2005watermill} and \cite{lafaye2008optimized}, the idea of query optimization for fingerprinting relational
databases subject to meeting usability constraints was proposed. The authors use declarative language to define usability constraints
for watermarking and fingerprinting relational databases. They optimize the watermark embedding by searching for
specific patterns such as aggregates and join computations. The idea of using query optimization for fingerprinting is interesting but the dependency of fingerprint decoding on the usability constraints may lead to incorrect fingerprint detection after an attack.

Fingerprinting techniques proposed by Guo et al. in \cite{guo2006fingerprinting} embed fingerprints in the database at two levels.
At the first level, the data is partitioned into $m$ partitions and $m$ length fingerprint is embedded in the data by embedding
 $i^{th}$ fingerprint bit in the selected tuples of $i^{th}$ partition. At the second level, the fingerprint is
extracted from the fingerprinted tuples and a confidence level is assigned to the extracted fingerprint. This fingerprint is used
as a secret key for the second level. The tuples already fingerprinted are not selected for fingerprint embedding at this level
to avoid conflict between the fingerprint embedding at both levels.


In \cite{zhou2007novel}, proposed a fingerprinting technique to identify a malicious traitor, who is redistributing digital content.  When a user access the database a fingerprinting process is invoked.
A manager module performs the fingerprinting task and a parameter manager module is used to store the
fingerprinting parameters. Authenticity of users is verified by the manager module; if the user is unauthentic, the fingerprinting process is executed. The encoding and decoding algorithm is the same as proposed in \cite{guo2006improved}.

Table \ref{tab:BRFTs} shows the comparison and summary of well known BRFTs based on different parameters shown in Figure \ref{fig:DBWMFramework}.

\begin{table*}[htbp]\tiny
  \centering
   \caption{Comparison and summary of well known BRFTs.}
     \label{tab:BRFTs}%
    \begin{tabular}{|l|l|l|l|l|l|l|l|l|l|l|}
       \hline
  \multicolumn{11}{|c|}{\textbf{Robust Techniques}}\\
    \hline
   \textbf{Sch.} & \textbf{MAT} &\textbf{ASM} & \textbf{TSM} & \textbf{GL} & \textbf{DMB} &\textbf{ECM} & \textbf{Dep.} & \textbf{Dist.}& \textbf{Opti.}&  \textbf{Rev.} \\
   \hline
    \cite{li2003constructing}& N & PRSG & PRSG & Bit & Blind & Voting & PK, Value & Yes & No &  No \\
    &  & based & based &  &  & & & && \\
    \hline
    \cite{li2005fingerprinting} & N     & PRSG & PRSG & Bit   & Blind & Voting & PK, Value & Yes& No &  No \\
          &       & based & based &       &       &       &       &  &&\\
          \hline
    \cite{liu2005block} & N     & All   & All   & Bit   & Yes   & -     & PK, Value & Yes & No &  No \\
    \hline
    \cite{guo2006fingerprinting} & N     & Arbitrary & SHF   & Bit   & Blind & -     & PK, Value & Yes & No &  No \\
          &       &       & based &       &       &       &       &  &&\\
          \hline
    \cite{zhou2007novel} & N     & Arbitrary & SHF   & Bit   & Blind & -     & PK, Value & Yes & No &  No \\
          &       &       & based &       &       &       &       &&&  \\
          \hline
    \cite{lafaye2008optimized} & N     & Arbitrary &PRSG & Bit   & Blind & Voting & PK, Value & Yes & No &  No \\
          &       &       & based &       &       &       &       & && \\
  \hline
    \end{tabular}%
\end{table*}%

\subsection{Data statistics-modifying techniques}

We further classify such techniques into two categories:
\begin{itemize}
  \item Bit pattern based data statistics-modifying techniques(BPBDSMT): In such techniques, a bit pattern is used as a watermark and these bits
are embedded into data statistics.
  \item Image based  data statistics-modifying techniques(IBDSMT): Image is used for watermark embedding in data statistics in such techniques.

\end{itemize}

\subsubsection{Bit pattern based data statistics-modifying techniques}
%

\paragraph{BPBDSMT with no specific watermark information}
The watermarking technique in \cite{sion2004rights} uses special marker tuples to virtually partition the data into
maximal number of subsets. The number of partitions should be equal to number of subsets. One watermark bit is
embedding into each subset subject to usability constraints. If the number of subsets are greater than the number of bits
in the watermark, an error correction mechanism (e.g., majority voting) is used to make the watermarking technique
resilient to the malicious attacks. The marker tuples are used in the watermark decoding stage for  data partitioning
and successful decoding of the watermark. Two arbitrarily chosen thresholds are used to decode the watermark bits.
This technique has two important shortcomings: (1) use of marker tuples for data partitioning makes the technique vulnerable
to synchronization errors during watermark decoding because the tuples deletion and insertion attacks on the position
of these tuples will change the boundaries of the partitions and hence the embedded bit in those partitions would
not be correctly detected; and (2) storing marker tuples for their use in decoding stage makes this technique not-blind.
Moreover, the thresholds, used for decoding an embedded bit, are chosen arbitrarily that also result in decoding errors. Another such technique as presented by Seb\'e et al. in \cite{seb2005noise} for numerical datasets while preserving the mean and variance of the watermarked attribute.


In
\cite{shehab2008watermarking}, the database is first partitioned into $m$ non-overlapping logical groups.
In this technique, bit encoding was modeled
as an optimization problem and then optimization algorithms were used to optimize one-bit watermark embedding subject to usability constraints specified by the data owner. Authors used a threshold
based approach and a majority voting scheme to decode the watermark.
This scheme is robust against tuple deletion, alteration and insertion attacks but it does not specify how to select attribute for watermarking. Deshpande et al. also proposed a similar watermarking technique in \cite{deshpande2009new}, that is a combination of \cite{sion2004rights} and
\cite{shehab2008watermarking}.

Huang et al. \cite{huang2009cluster} presented a cluster based watermarking algorithm for relational databases.
During the embedding phase, the data is first clustered into $k$ clusters using k-means clustering algorithm \cite{lloyd1982least}. Such techniques are suspectable to alteration attacks if an attribute value is modified by the attacker to an extent where the cluster  of a attribute is changed.


Kamran and Farooq coined the idea of information-preserving watermarking in \cite{kamran2011information}. This technique is focused on watermarking of
relational databases containing patient records ( such databases are also termed as electronic health records
(EHR) or electronic medical records (EMR)). Authors first identify the numeric attributes for watermarking using information gain as a measure. Then they select attributes with low
information gain for watermarking such that change in their values does not have any impact on the diagnosis prediction by the data mining
algorithms. Authors also specify equations for calculating the bounds for changes in the numeric features. The change is then optimized and embedded in the selected attributes in every row
of the EMR subject to
usability constraints. Watermark decoding also utilizes the knowledge of information gain of the attributes. A watermark decoder is used to decode
watermark bits from each row of the marked attributes. A majority voting scheme is applied to combat decoding errors. Another interesting feature
of this technique is that it does not require any secret key for watermark embedding or decoding yet this scheme is resilient to tuple insertion, deletion
and alteration attacks because watermark is embedded in each row of the table.

For minimizing (or controlling) data distortions, Kamran et al. in \cite{kamran2013robust} limit the to-be-watermarked tuples by suing a threshold and even hash values; while for high robustness is achieved through once-for-all usability constraints. Similarly, In \cite{iftikhar2015rrw}, Saman et al. extended the work of \cite{kamran2011information} to come up with a reversible technique while considering mutual information \emph{MI} instead of information gain \emph{IG} for attribute selection. Other such techniques are presented in \cite{franco2014robust} and \cite{franco2015robust} where authors consider the histogram modulation (in \cite{franco2014robust}) and semantic properties (in \cite{franco2015robust}) of numeric attributes during watermark embedding to control the data distortions.

Table \ref{tab:BPBDSMTsMeaningLess} shows the comparison and summary of well known BPBDSMTs with no specific watermark information based on different parameters shown in Figure \ref{fig:DBWMFramework}.

\begin{table*}[htbp]\tiny
  \centering
   \caption{Comparison and summary of well known BPBDSMTs with no specific watermark information.}
     \label{tab:BPBDSMTsMeaningLess}%
    \begin{tabular}{|l|l|l|l|l|l|l|l|l|l|l|}
       \hline
  \multicolumn{11}{|c|}{\textbf{Robust Techniques}}\\
    \hline
   \textbf{Sch.} & \textbf{MAT} &\textbf{ASM} & \textbf{TSM} & \textbf{GL} & \textbf{DMB} &\textbf{ECM} & \textbf{Dep.} & \textbf{Dist.}& \textbf{Opti.}&  \textbf{Rev.} \\
   \hline
    \cite{sion2004rights} & N & Arbitrary & SHF   & Value & Semi Blind & Voting & PK, &  Yes& No &  No \\
    & & & based & & & & Markers &  &&\\
       \hline
    \cite{seb2005noise} & N     & Arbitrary & Arbitrary & Value & Semi Blind & -     & Value & Yes & No &  No \\
       \hline
    \cite{shehab2008watermarking} & N     & Arbitrary & All   & Value & Blind & Voting & PK    & Yes & Yes &  No \\
       \hline
    \cite{kamran2011information} & N     & IG    & All   & Value & Semi Blind & Voting & IG    & Yes & Yes &  No \\
     \hline
     \cite{kamran2013robust} & N     & Arbitrary    & SHF   & Value & Blind & Voting & Value    & Yes & No &  No \\
     \hline
     \cite{franco2014robust} & N     & Arbitrary    & SHF   & Value & Blind & Voting & Value    & Yes & No &  Yes \\
     \hline
     \cite{iftikhar2015rrw} & N     & MI    & All   & Value & Semi Blind & Voting & MI    & Yes & Yes &  Yes \\
     \hline
     \cite{franco2015robust} & N     & Arbitrary    & SHF   & Value & Blind & - & Value    & Yes & No &  No \\
     \hline

    \end{tabular}%
\end{table*}%

\paragraph{BPBDSMT with specific watermark information}

In \cite{zhang2005method}, Zhang et al. proposed  a watermarking scheme using cloud models.
They form the cloud model using different parameter and later watermark the database using these parameters. In the detection phase, a backward cloud generation algorithm is used
to extract the embedded watermark. Finally, a cloud
algorithm is used for matching clouds of watermark embedding and decoding. The watermark decoding is not blind
because the original database is required during the decoding phase.

In \cite{zhang2006reversible}, Zhang et al. proposed a histogram expansion based reversible watermarking scheme
for relational databases. For watermark embedding, authors use the histogram expansion techniques employing overhead information. The overhead information is used to distinguish the
original digits and the watermarked digits. Authors use the Haar wavelet transform for watermarked database
attribute identification. For embedded decoding, the inverse of watermarking process is applied. Such techniques need to keep a track of overhead information.

Fu et al. \cite{fu2007novel} used spread spectrum for selecting database tuples for watermarking.
A function $W(t,k)$ is also used to yield
the watermark signal $W= {+1,-1}$ by using direct sequence spread spectrum and then this signal is embedded in the data as watermark. In the watermark decoding phase, tuples are grouped, attributes are sorted and watermark is detected using even parity check. Finally, majority voting scheme is applied to eventually decode the watermark bits. This technique is vulnerable to alteration attacks that may alter the values of marked attributes.


In \cite{jiang2009watermarking} a discrete wavelet transform (DWT) \cite{shensa1992discrete} based watermarking technique for relational
databases was proposed.
A data $X$ is watermarked as $C^{W}=C+\alpha \phi (\beta)\Omega$, where $C$ is the original carrier, $C^{W}$ is the carrier after watermark embedding, $\Omega$ is the watermark template sequence, $\alpha$ is the intensive factor and $L$ is the length of watermark string. $\Omega$ is generated, by a secret key, as a pseudo random sequence. This scheme suffers form lack of robustness as embedding of watermark in only high frequency coefficients make it easy for the attacker to target one particular group of tuples to corrupt the watermark.


Table \ref{tab:BPBDSMTsMeaningFull} shows the comparison and summary of well known BPBDSMTs with specific watermark information based on different parameters shown in Figure \ref{fig:DBWMFramework}.

\begin{table*}[htbp]\tiny
  \centering
   \caption{Comparison and summary of well known BPBDSMTs with specific watermark information.}
     \label{tab:BPBDSMTsMeaningFull}%
    \begin{tabular}{|l|l|l|l|l|l|l|l|l|l|l|}
       \hline
  \multicolumn{11}{|c|}{\textbf{Robust Techniques}}\\
    \hline
   \textbf{Sch.} & \textbf{MAT} &\textbf{ASM} & \textbf{TSM} & \textbf{GL} & \textbf{DMB} &\textbf{ECM} & \textbf{Dep.} & \textbf{Dist.}& \textbf{Opti.}&  \textbf{Rev.} \\
   \hline
    \cite{zhang2005method} & N & Arbitrary & Arbitrary & Value & Not & -     & PK & Yes & No &  No \\
     & & & & & Blind &       &  &  &&\\
     \hline
    \cite{zhang2006reversible} & N     & Arbitrary & Arbitrary & Value & Blind & -     & PK, Value & Yes & No &  Yes \\
    \hline
    \cite{fu2007novel} & N     & Arbitrary & SHF   & Value & Blind & EPC, Voting & PK    & Yes& No &  No \\
          &       &       & based &       &       &       &       &  &&\\
          \hline
           \cite{farfoura2012blind} & N     & SHF   & SHF   & Value & Blind & Voting & PK, Value & Yes & No &  No \\
          &       & based & based &       &       &       &       & && \\
          \hline
           \end{tabular}%
\end{table*}%

\subsubsection{Image based data statistics-modifying techniques}
%

In \cite{zhang2004watermarking}, Zhang et al. first divided into logical groups. Each group has its size equal to the size
of an image. The pixel values $(v_{0},v_{1},v_{2},.....,v_{m})$ are embedded into the logical groups of the data. A pixel value $v_i{}$ is embedded in such a way that if $v_{i}=255$, the attribute value modulo 3 is set to 1 and if $v_{i}=0$, the attribute value modulo 3 is set to 2; otherwise, the pixel value is added to the attribute value. In the watermark detection phase, the attribute value modulo 3 is computed. If it returns 1 then $v_{i}=255$ is detected and if $v_{i}=0$ then 2 is detected; otherwise, a pixel value $v_{i}$ is calculated by subtracting the integer part of
the attribute value $A_{i}$ from an overall attribute value $A_{i}$. The result is multiplied by 255 yielding the value of $v_{i}$, and hence watermark is decoded. If the attribute value is changed by an attacker, the watermark decoding accuracy is degraded.

In \cite{tsai2007fragile} Tsai et al. also use image for fragile watermarking of database using support vector regression (SVR). The purpose of such techniques is to detect malicious database tampering.


In \cite{odeh2008watermarking} Odeh and Al-Haj proposed a watermarking scheme for relational databases using an image. In watermark embedding phase, an image is converted to bit strings, bits are grouped with each group having size of 5 bits. The bits are then converted to a decimal number. A time attribute is selected for watermark embedding. For this purpose, seconds portion of the time attribute is marked. In watermarking detection phase, the marked tuples are identified by using secret key, and the appended decimal number is extracted from seconds part of time attribute. The number is converted to bits, and bits are grouped in a group of 5 bits. Image is again constructed using these bits. For experiments, the authors have created their own database with multiple time attributes but in practice most databases usually do not have a great number of such attributes.

In \cite{farfoura2010novel}, an image is first converted into a stream of watermark bits consisting of 0's, and 1's.
The watermark is embedded in the fractional part of the numeric attribute. A hash function is applied to select the tuples for watermarking. For a user $ID$, a database $DB_{name}$, with version $version$, the secret key $K$ is computed using the relation: $K = H(ID||DB_{name}||version|...)$. Authors assume two intensities for each pixel, the original intensity and the predicted intensity and use their difference to encode watermark.

Table \ref{tab:IBDSMTs} shows the comparison and summary of well known IBDSMTs based on different parameters shown in Figure \ref{fig:DBWMFramework}.

\begin{table*}[htbp]\tiny
  \centering
   \caption{Comparison and summary of well known IBDSMTs.}
     \label{tab:IBDSMTs}%
   \begin{tabular}{|l|l|l|l|l|l|l|l|l|l|l|}
       \hline
  \multicolumn{11}{|c|}{\textbf{Robust Techniques}}\\
    \hline
   \textbf{Sch.} & \textbf{MAT} &\textbf{ASM} & \textbf{TSM} & \textbf{GL} & \textbf{DMB} &\textbf{ECM} & \textbf{Dep.} & \textbf{Dist.}& \textbf{Opti.}&  \textbf{Rev.} \\
   \hline
   \cite{zhang2004watermarking} & N & Arbitrary & All   & Value &Blind & - & PK & Yes & No &  No \\
   \hline
    \cite{farfoura2010novel} & N     & SHF   & SHF   & Value & Blind & Voting & PK    & Yes & No &  No \\
          &       & based & based &       &       &       &       &  &&\\
   \hline
    \end{tabular}%
\end{table*}%

%

\subsection{Constrained Data Content-Modifying Techniques}

We further classify such techniques into two categories:
\begin{itemize}
  \item Tuple based constrained data content-modifying techniques(TBCDCMT): Such techniques modify database contents modification at the tuple level.
  \item Attribute based constrained data content-modifying techniques(ABCDCMT): Such techniques involve modification of database content at an attribute level.
  \item An image based  constrained data content-modifying techniques(IBCDCMT): Image is used, for watermark embedding in such
  techniques.

\end{itemize}

\subsubsection{Tuple based constrained data content-modifying techniques}
%

\paragraph{TBCDCMT with no specific watermark information}

Gross-Amblard \cite{gross2003query} proposed a query-preserving watermarking of XML documents and relational databases. 
For database watermarking, the author used Gaifman graph\footnote{``A Gaifman graph is
the graph in which nodes are the variables of the problem and an edge joins a pair of variables if the two variables occur together in a
constraint'' \cite{rossi2006handbook}.} while preserving the database local queries.  These queries are expressed in plain SQL for database operations like adding grouping and aggregate
functions. The database tuples are marked in such a way that the query result remains unaltered. Such techniques, however, are vulnerable to tuple insertion, deletion and alteration attacks because the values of the marked tuple may be disturbed by these attacks.


Li et al. \cite{li2004tamper} proposed a watermarking scheme for detecting alterations in the database relation
containing categorical data based on the tuples order.  For increasing the watermark embedding capacity in such algorithms, authors suggest to use Myrvold and Ruskeys linear permutation unranking algorithm \cite{myrvold2001ranking} for exchanging the position
of tuples for distortion free watermarking \cite{li2008database}. Such schemes are aimed at watermark fragility; therefore, they can not be used for ownership protection. Another distortion-free Fragile Watermarking technique in \cite{bhattacharya2009distortion} works for categorical data.  A similar technique is proposed in \cite{arun2012distortion} and it uses tuple pairs for fragile distortion-free watermarking. Other such techniques have been presented in \cite{guo2011fragile} and \cite{li2011asymmetric}.


Kamel \cite{kamel2009schema} proposed the idea of using R-tree for watermarking relational databases to protect the integrity of databases. In this technique R-tree is used to represent entries in the relation as minimum bounding rectangle ($MBR$). In the watermark encoding phase, a
one-to-one mapping function between all permutations of the entries and the numeric value of watermark is performed. A \emph{factorial number system} is used to convert the decimal value of watermark to factorial form. Watermark embedding starts by sorting the entries in an R-tree node on the basis of a reference order $E_{k}$. The sorting order may be on the basis of: (1) $x$ value at the $MBR$; (2) the area of $MBR$; and (3) the perimeter of $MBR$. A circular-left-shift operation is applied on the selected entries of R-tree for mark encoding. The integrity of databases is checked by verifying the order of $MBR$ coordinates (reference orders). If an attack aims at modifying this order, the watermark decoding accuracy is affected. In \cite{kamel2011toward}, Kamel et al. used the idea of a $sensitive$ attribute for tuple ordering.



Hamadou et al. proposed a zero-watermarking technique in \cite{Hamadou2011fragile} to counter additive watermarking
attacks. The technique is named zero-watermarking because watermark is not actually inserted in the database instead its information
is registered with the certification authority ($CA$). Another technique to counter additive attack is presented in \cite{chen2008self}.

Table \ref{tab:TBCDCsMeaningTLess} shows the comparison and summary of well known TBCDCMTs with no specific watermark information based on different parameters shown in Figure \ref{fig:DBWMFramework}.

\begin{table*}[htbp]\tiny
  \centering
   \caption{Comparison and summary of well known TBCDCMTs with no specific watermark information.}
     \label{tab:TBCDCsMeaningTLess}%
    \begin{tabular}{|l|l|l|l|l|l|l|l|l|l|l|}
       \hline
  \multicolumn{11}{|c|}{\textbf{Robust Techniques}}\\
    \hline
   \textbf{Sch.} & \textbf{MAT} &\textbf{ASM} & \textbf{TSM} & \textbf{GL} & \textbf{DMB} &\textbf{ECM} & \textbf{Dep.} & \textbf{Dist.}& \textbf{Opti.}&  \textbf{Rev.} \\
    \hline
  \multicolumn{11}{|c|}{\textbf{Fragile Techniques}}\\
   \hline
   \textbf{Sch.} & \textbf{MAT} &\textbf{ASM} & \textbf{TSM} & \textbf{Loc.} & \textbf{DMB} &\textbf{ECM} & \textbf{Dep.} & \textbf{Dist.}& \textbf{Opti.}&  \textbf{Char.} \\
   \hline
    \cite{li2004tamper} & C     & Arbitrary & SHF   & Tuple & Blind & -     & PK, Order & No & No & No \\ 
          &       &       & based &       &       &       &       & && \\
            \hline
   \cite{kamel2009schema} & All     & Index & Index & - & Blind & -     & Reference & No & No & No \\ 
          &       & based & based &  &       &       & order &  &&\\
            \hline
    \end{tabular}%
\end{table*}%

\paragraph{TBCDCMT with specific watermark information}

A publicly verifiable watermarking scheme was proposed by Li et. al in \cite{li2006publicly}.
This technique works for watermarking attributes with any data type including numeric, strings or boolean.
The \emph{watermark key} is a public key
that is generated by applying one-way hash functions on personal information (e.g., owner's identity) of the author and
the database information (e.g., database name and version). The watermark $W$ is generated using this \emph{secret key}.
The watermark is also a database relation and has the same number of attributes and tuples as in the original database relation $R$. The number of watermark bits are decided by using a \emph{watermark generation parameter}. A pseudo random sequence generator is used for watermark embedding. This scheme can detect
any modifications made to MSBs but changes made in other bits cannot be detected using this technique.

%

Bhattacharya et al. \cite{bhattacharya2009generic} proposed a modified version of watermarking technique presented in  \cite{li2006publicly} by using a private secret key instead of a public key. The technique performs the data partitioning and extracts a binary watermark as an image for proving the data ownership.


Other such techniques are presented in \cite{el2009new}, and \cite{el2010novel} but they are slight modifications of the techniques discussed above.

Table \ref{tab:TBCDCMTsMeaningFull} shows the summary and comparison of well known TBCDCMTs with specific watermark information based on different parameters shown in Figure \ref{fig:DBWMFramework}.

\begin{table*}[htbp]\tiny
  \centering
   \caption{Summary and comparison of well known TBCDCMTs with specific watermark information.}
     \label{tab:TBCDCMTsMeaningFull}%
    \begin{tabular}{|l|l|l|l|l|l|l|l|l|l|l|}
       \hline
  \multicolumn{11}{|c|}{\textbf{Robust Techniques}}\\
    \hline
   \textbf{Sch.} & \textbf{MAT} &\textbf{ASM} & \textbf{TSM} & \textbf{GL} & \textbf{DMB} &\textbf{ECM} & \textbf{Dep.} & \textbf{Dist.}& \textbf{Opti.}&  \textbf{Rev.} \\
   \hline
    \cite{li2006publicly} & All     & All   & All   & Table & Blind & -     & PK, Value & No & No &  No \\
    \hline
    \cite{bhattacharya2009generic} & All     & All   & All   & Table & Blind & -     & PK, Value & No & No &  No \\
    \hline
    \end{tabular}%
  \label{tab:addlabel}%
\end{table*}%

\subsubsection{Attribute based constrained data content-modifying techniques}
%

\paragraph{ABCDCMT with no specific watermark information}

In \cite{halder2010persistent} Halder et al. introduced the notion of persistent watermarking of relational databases. For this purpose, this
technique identifies two type of features from the database: (1)  Stable cells are the database cells that are not affected by any set of
queries $Q$ (Select, Insert, Update, and Delete) at all; and (2) Semantic properties.
The stable part and semantic properties are watermarked separately. In watermark embedding phase, database is first partitioned into $m$
non-overlapping partitions. The watermark is converted into a bit string of length $m$. The stable part and semantic properties are
watermarked such that the usability constraints are satisfied. In the watermark decoding phase, data partitioning is performed and embedded bits are detected from semantic properties and the stable cell. Only the tuples having same primary keys in watermarked and suspected databases are used for watermark detection.
In \cite{halder2011persistent} Halder et al. extend their work by embedding both the private and the public watermark.
 The purpose of private watermarking is proving the ownership of the data and the public key is used to check the authenticity of data without worrying for revealing secret parameters.

In \cite{shah2011query} a query preserving watermarking was been proposed. This technique selects alphanumeric attributes for watermarking and alters the case of the attribute value such that the results of a query are preserved. A secret key is used to select tuple and alphanumeric attribute for watermark embedding. The letters of selected attribute values are changed to a sentence case if a watermark bit is 0; otherwise, it the value is converted to title case. The watermark detection is the inverse process of watermark embedding. This technique is not resilient against tuple alteration attacks because the attacker can change the case of the attribute values without degrading the data quality.

In  \cite{kamran2013formal}, a distortion-free watermarking technique has been proposed that works on all type of attributes and also considers the importance of attributes for different application domains. The watermark encoding and decoding is based on secret ordering for non-numeric attributes while for numeric-feature the technique of \cite{kamran2011information} is extended by using a secret threshold for the information gain \emph{IG} of the attributes. Other such techniques are presented in \cite{hanyurwimfura2010text}, \cite{uzun2008security}, and \cite{el2010novel} but for brevity (and to adhere to the page limit of the Journal), they are not discussed here.

Table \ref{tab:ABCDCMTs} shows the comparison and summary of well known ABCDCMTs with no specific watermark information based on different parameters shown in Figure \ref{fig:DBWMFramework}.

\begin{table*}[htbp]\tiny
  \centering
   \caption{Comparison and summary of well known ABCDCMTs with no specific watermark information.}
     \label{tab:ABCDCMTs}%
    \begin{tabular}{|l|l|l|l|l|l|l|l|l|l|l|}
       \hline
  \multicolumn{11}{|c|}{\textbf{Robust Techniques}}\\
    \hline
   \textbf{Sch.} & \textbf{MAT} &\textbf{ASM} & \textbf{TSM} & \textbf{GL} & \textbf{DMB} &\textbf{ECM} & \textbf{Dep.} & \textbf{Dist.}& \textbf{Opti.}&  \textbf{Rev.} \\
  \hline
     \cite{kamran2013formal} & All    & IG   & All   & Value & Blind & Voting     & IG    & No & Yes & No \\
     \hline
  \multicolumn{11}{|c|}{\textbf{Fragile Techniques}}\\
   \hline
   \textbf{Sch.} & \textbf{MAT} &\textbf{ASM} & \textbf{TSM} & \textbf{Loc.} & \textbf{DMB} &\textbf{ECM} & \textbf{Dep.} & \textbf{Dist.}& \textbf{Opti.}&  \textbf{Char.} \\
   \hline
    \cite{halder2011persistent} & N & Property & PRSG & - & Blind & Voting & PK & No & No & No \\ 
    &       &  & based & &  &  & & && \\
     \hline
   \cite{shah2011query} & AN    & SHF   & SHF   & - & Blind & -     & PK    & No & No & No \\ 
          &       & based & based &       &       &       &       & && \\
     \hline
    \end{tabular}%
\end{table*}%

\paragraph{ABCDCMT with specific watermark information}

The techniques like  \cite{gamal2008simple}, \cite{prasannakumari2009robust} are based on addition of a new column in the database relation.
The values in the new column are calculated by aggregating the numeric columns present in a database relation. The authors suggest
to lock this column using a secret key to ensure watermark security. But they do not clearly highlight the locking mechanism.
The major drawback of such techniques is that the watermark information is lost by dropping only one attribute from the database relation without loss of data usability.

Bedi et al. \cite{bedi2011new} proposed a watermarking approach for non-numeric data. They first generate a secret key from eigen values of \emph{tuple-Relation} matrix for a tuple. A low impact attribute (e.g., address) is selected for watermark embedding. In another technique, Bedi et al. \cite{rajneesh2011unique} use abbreviations of the words for watermarking. Similarly, in \cite{khanduja2012robust}, vowels are used to embed the watermark. Such techniques are vulnerable to alteration attacks and even benign updates (changing of values of marked attributes) might result in watermark loss without significantly degrading the data quality.

Another such technique for medical data is proposed in \cite{coatrieux2011lossless}. This technique is based on shifting of attributes based on the histogram constructed from the tuple values. This technique works with categorical attributes and is aimed to ensure data integrity.

The technique in \cite{khan2013fragile} uses numeric attributes to generate and verify a distortion-free fragile watermark for detecting and characterizing the malicious modifications in the database relations. For numerical attibutes, authors use digit frequency, length and range of data values for watermark generation and verification. This technique is also able to characterize the malicious attacks (or modifications).

Table \ref{tab:ABCDCMT} summarizes the well known ABCDCMTs.

\begin{table*}[htbp]\tiny
  \centering

  \caption{Comparison and summary of well known ABCDCMTs.}
   \label{tab:ABCDCMT}%
    \begin{tabular}{|l|l|l|l|l|l|l|l|l|l|l|}
       \hline
  \multicolumn{11}{|c|}{\textbf{Fragile Techniques}}\\
   \hline
   \textbf{Sch.} & \textbf{MAT} &\textbf{ASM} & \textbf{TSM} & \textbf{Loc.} & \textbf{DMB} &\textbf{ECM} & \textbf{Dep.} & \textbf{Dist.}& \textbf{Opti.}&  \textbf{Char.} \\
   \hline
    \cite{khan2013fragile} & N     & Property & All&-& Blind & - & Value & No& No & Yes \\ 
     \hline
    \end{tabular}%

\end{table*}%


\subsubsection{Image based constrained data content-modifying techniques}
%

In \cite{tsai2006database}, a table $T$ with attributes $(P,A_{1},A_{2},.....,A_{j})$ is watermarked to detect tampering with the database. A watermark $WM$ of length $N$ is computed from an image. An image $SD$ is also generated from the certification verification code to use it for watermark detection. For the certification code verification, watermark detection process starts by computing  certification codes by applying public key on the image $SD$. All features are combined to yield a vector $C'$. Finally, an $XOR$ operation is applied on $C'$ and certification code to compute the watermark $WM'$. If it matches with embedded watermark then database integrity is proved otherwise it is tampered.




In \cite{al2008robust}, Al-Haj et al. proposed a watermarking scheme for non-numeric databases based on insertion of a double-space in multi-word attribute. This technique is not resilient to alteration attack because an attacker can easily remove double spaces from the database without affecting the data usability. Moreover, benign updates also corrupt the embedded watermark for such techniques.

In \cite{Bhattacharya2010Database}, database is first partitioned and a pseudo random sequence generator is used to select an attribute for
watermarking. This technique does not reset any bit of the original data instead it uses $m$ most significant bits (MSBs) and $n$ LSBs  of the
selected field to generate the watermark for that field such that $m+n=8$.  The watermark for a particular partition is converted to a gray scale image, so the value of each cell ranges between 0 and 255. The watermark does not bring any distortion in the database. In the watermark detection process, the database is again partitioned and the watermark is generated using the same procedure as in watermark embedding. The watermark bits are matched. If the match count is equal to $\omega$ (with $\omega= rows \times columns$), the watermark is successfully decoded. A zero-distortion authentication watermarking \cite{wu2003zero} is used for authenticating the table without any distortions. This technique is only useful for proving integrity of data.

Table \ref{tab:IBCDCMT} shows the comparison and summary of well known IBCDCMTs based on different parameters shown in Figure \ref{fig:DBWMFramework}.

\begin{table*}[htbp]\tiny
  \centering

  \caption{Comparison and summary of well known IBCDCMTs.}
   \label{tab:IBCDCMT}%
    \begin{tabular}{|l|l|l|l|l|l|l|l|l|l|l|}
       \hline
  \multicolumn{11}{|c|}{\textbf{Fragile Techniques}}\\
   \hline
   \textbf{Sch.} & \textbf{MAT} &\textbf{ASM} & \textbf{TSM} & \textbf{Loc.} & \textbf{DMB} &\textbf{ECM} & \textbf{Dep.} & \textbf{Dist.}& \textbf{Opti.}&  \textbf{Char.} \\
   \hline
    \cite{tsai2006database} & All     & All & All&Tuple& Blind & - & PK & Yes & No & No \\ 
     \hline
    \cite{Bhattacharya2010Database} & All & PRSG & All & -& Blind & -     & PK    & No & No & No \\ 
         &       &       based &  & &       &       &       &  &&\\
  \hline
    \end{tabular}%

\end{table*}%

%

\section{Applications, Comparison and Future Direction for Different Techniques}
\label{sec:ComparisonofTechniques}
In this section, we compare the three classes -- BRT, DSMT, and CDCMT -- of techniques using two contexts: (1) robustness against malicious attacks;
and (2) data usability (again there parameters were short-listed from different parameters listed in Figure \ref{fig:DBWMFramework}). In the first context, we report the resilience of the watermarking/fingerprinting techniques against malicious attacks
that have been discussed in section \ref{sec:Introduction}. In the data usability context, we report methods of ensuring data quality (used in
the watermarking/fingerprinting techniques) and depict scenarios where usability is a serious concern.
The robustness of each type of techniques is depicted in Table \ref{table:RobustnessAgainstMaliciousAttacks}. In this table $R_{1}$, $R_{2}$, $R_{3}$,
$R_{4}$, and $R_{5}$ denote the
robustness of watermarking techniques against $A_{1}$, $A_{2}$, $A_{3}$, $A_{4}$, and $A_{5}$ attacks respectively.

\begin{table*}[htbp]\scriptsize
  \centering
   \caption{Robustness of watermarking techniques against malicious attacks}
  \label{table:RobustnessAgainstMaliciousAttacks}
    \begin{tabular}{|c|l|l|l|l|l|}
   \hline
   \textbf{Class of Techniques} & \textbf{$R_{1}$ }   & \textbf{$R_{2}$}    & \textbf{$R_{3}$}    & \textbf{$R_{4}$}    & \textbf{$R_{5}$} \\
    \hline
     & Usually resilient except&Usually resilient except&Usually not resilient&Usually not resilient& Usually not resilient \\
    &techniques that&techniques that:\ &because small&due to facts depicted&apart from techniques \\
    {\textbf{BRT}}&use specific order&use specific order&manipulations can easily&in $R_{1}$, $R_{2}$, and $R_{3}$.&that are specifically \\
    &(or position) of tuples&(or position) of tuples& disturb the marked bits& &designed for tackling\\
    &for watermark&for watermark insertion.&and even multi-bit&&attacks of type $A_{5}$\\
     &insertion.&&watermarks can be corrupted.&&and they still have\\
     &&&&&limitations.\\
    \hline
     & Usually resilient except&Usually resilient except&Usually resilient.&Usually techniques& Usually not resilient \\
    &techniques that&techniques that&&that mark more tuples &apart from techniques \\
    &use specific order&use specific order&&are more robust to &that are specifically \\
    &(or position) of tuples&(or position) of tuples& &such attacks. But still&designed for tackling\\
   {\textbf{DSMT}} &for watermark&for watermark insertion.&&this type of attack&attacks of type $A_{5}$\\
     &insertion.&&&needs to be studied &and they still have\\
      &&&&in depth because most of&limitations.\\
        &&&&the techniques proposed&\\
         &&&&so far do not&\\
         &&&&discuss this attack.&\\
    \hline
     & Usually resilient except&Usually not resilient&Usually not resilient&Usually not resilient& Usually not resilient \\
    &techniques that& because deletion attack&because small&due to the facts depicted&apart from techniques \\
    &use specific order&may cause deletion of&manipulations can easily&in $R_{1}$, $R_{2}$, and $R_{3}$.&that are specifically \\
    {\textbf{CDCMT}}&(or position) of tuples&newly inserted content &disturb the marked bits&&designed for tackling\\
    &or tuples for&if new content is inserted&without degrading the&&attacks of type $A_{5}$\\
     &watermark insertion.&during watermark&data usability significantly.&&and they still have\\
      &&embedding.&&&limitations.\\
       \hline

    \end{tabular}%
\end{table*}%

As depicted in table \ref{table:RobustnessAgainstMaliciousAttacks}, the DSMT techniques are relatively better against most type of attacks. ut attacks of type $A_{4}$ and $A_{5}$ still expose vulnerabilities of these techniques. In $A_{4}$ type attacks of
the attacker launches a combination of the $A_{1}$, $A_{2}$, $A_{3}$ attacks. Kamran and Farooq address this attack in \cite{kamran2011information}
but most of other techniques do not consider these type of attacks in their robustness study. \emph{However, most of the techniques do not consider these type of attacks in their robustness study; therefore, next generation watermarking techniques should incorporate mechanisms that counter $A_{4}$ attacks}.

In $A_{5}$ type attack, Mallory claims false ownership of the data by inserting his watermark into Alice's watermarked data. The techniques presented in \cite{zhou2007additive}, \cite{manjula2010new},  \cite{chen2008self}, \cite{Hamadou2011fragile}, \cite{li2004defending},
\cite{gupta2009database} provide mechanisms to counter such attacks, but these techniques require the role of third party with assumptions like:
(1) an attacker will not change the name of the attributes in a database relation; (2) an attacker would not want to disturb the Alice's
watermark to combat data errors;
and (3) the additive watermark will significantly degrade the data quality; as a consequence, Alice does not need to claim the ownership of this type of compromised data. It is important to emphasize that Mallory can surpass these assumptions while mantaining, for example, simply
changing the names of attributes does not affect the data quality. Moreover, Mallory may also want to corrupt the Alice's watermark
during $A_{5}$ type attack. Additionally, data has intrinsic value; therefore, it is not a great idea for Alice to withheld ownership claim of the data.
\emph{So, $A_{5}$ type of attacks also pose a new challenge to researchers of this field}. Such challenges are also valid for fingerprinting techniques because they also require the watermark to be robust.

The fragile watermarking techniques provide database integrity and tamper proofing in various application domains because they are sensitive to low intensity malicious attacks. But in some situations it can favor the attacking scenarios in which Mallory can safely claim the ownership of databases watermarked by such techniques by attacking a small part of the marked database. So, we think this also pose a unique challenge to the researchers of this field to come with fragile watermarking techniques that can also provide the mechanism for ownership protection for more generic applications that might require both tamperproofing and ownership protection.

In order to emphasize the problem of defining usability constraints, Table
\ref{table:Usability} shows the usability constraints defining method and applicability of BRT, DSMT, and CDCMT type of techniques.
All techniques require an owner to manually define the usability constraints and this process is application dependent; therefore an owner might need to define different usability constraints for each type of intended use of the data.  This method is cumbersome  because the data owner has to model a different set of usability constraints after mutual consensus with the intended
user. Moreover, he will have to watermark a given database for every type of user.  This activity might expose the watermarked rows to an attacker because he may easily compare the rows of two datasets and
guess the watermarked rows by calculating the difference between the corresponding rows of two datasets. \emph{Therefore, there is a need of a model
that automatically defines the data usability constraints such that the data quality is ensured and the data owner is not required to waste his resources to come up with an agreed (after mutual consensus with the intended
data user) data usability constraints}. This model should ideally be application independent; therefore, it is used for proving data ownership,
 tamper proofing, and data integrity, and for other applications of watermarking and fingerprinting. This model should encompass the
 intelligence to assure the data quality of the watermarked database for its potential use in every type of intended application. Moreover, the significance of a meaningful watermark (for instance, a biometric feature like voice, image and so on) can also be tested and emphasized.

\begin{table*}[htbp]\scriptsize
  \centering
   \caption{Usability constraints defining method and applicability of watermarking techniques}
  \label{table:Usability}
    \begin{tabular}{|c|l|l|}
   \hline
   \textbf{Class of Techniques} & \textbf{Usability constraints defining method}
   & \textbf{Applicability}\\
    \hline
   \textbf{BRT} &Application dependent, defined by the mutual&Proving data ownership. Fragile\\
   &consensus of the data owner and the data user.& techniques can be used for temper detection.\\
    \hline
   \textbf{DSMT}&Application dependent, defined by the mutual&Proving data ownership. Fragile\\
    &consensus of the data owner and the data user.&techniques can be used for temper detection.\\
    \hline
   \textbf{CDCMT}
&Application dependent, defined by the mutual&Proving data ownership\\
&consensus of the data owner and the data user.&and data integrity.\\
&&Also, protect against tampering.\\
\hline

    \end{tabular}%
\end{table*}%

\section{Conclusion}
\label{sec:Conclusion}
In this paper, a comprehensive review of the database watermarking and fingerprinting techniques, proposed to date, has been presented.
The classification of techniques has been done on the basis of watermarking technique and the orientation of the inserted watermark. Using this nomenclature, three types of top level classes are presented: BRT, DSMT, and CDCMT. A comparison of these techniques -- on the basis of robustness against malicious attacks, and the method for defining usability constraints
defining method -- has also been given. The DSMT techniques appear to be the best among different
available alternatives, but they do have certain limitations in different contexts. We have also pointed out the directions for future
research in this and related areas. To conclude, intelligent watermarking technique that ensure data quality and also are efficient and effective against malicious attacks is an important requirement.
%
%

%
\begin{IEEEbiography}[{\includegraphics[width=1in,height=1.25in,clip,keepaspectratio]{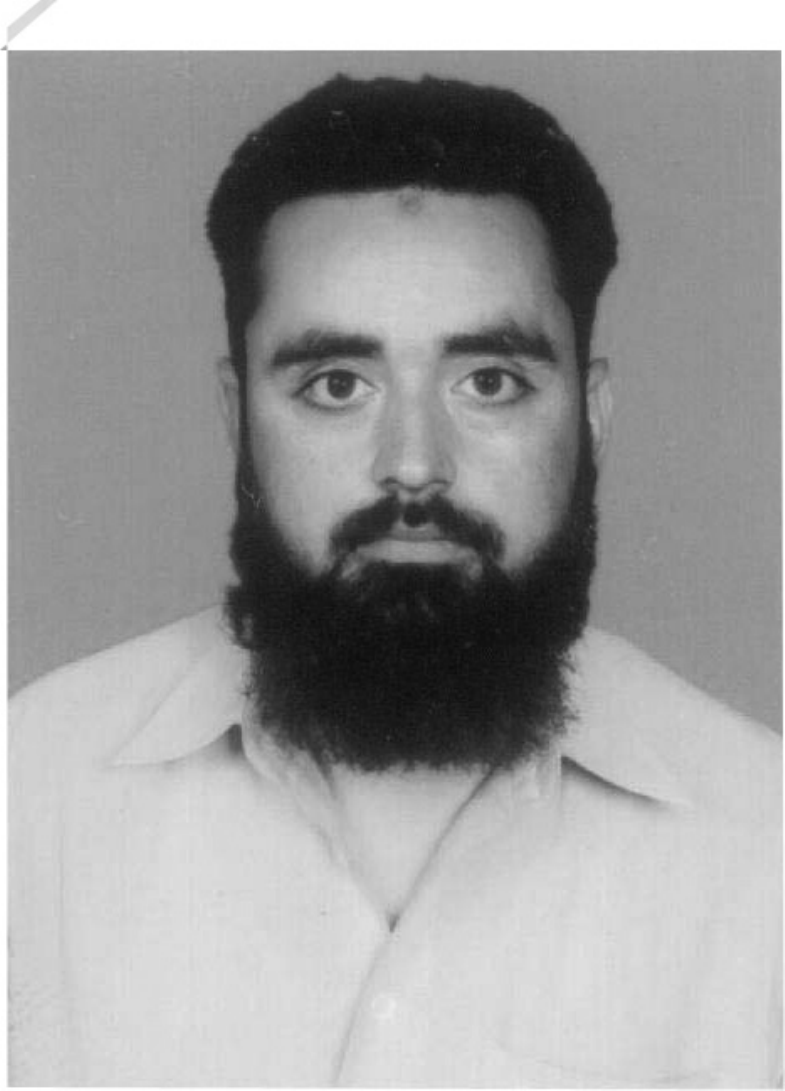}}]{M. Kamran}
 got his BS and MS degree in Computer Science in 2005 and
2008 respectively. Currently he is working as Assistant Professor of Computer Science at COMSATS Institute of Information Technology Wah Campus, Wah Cantt,
Pakistan. His research interests include the use of machine
learning, evolutionary computations, big data analytic, data security, and decision support systems.
\end{IEEEbiography}
\vspace*{-2\baselineskip}
\begin{IEEEbiography}[{\includegraphics[width=1in,height=1.25in,clip,keepaspectratio]{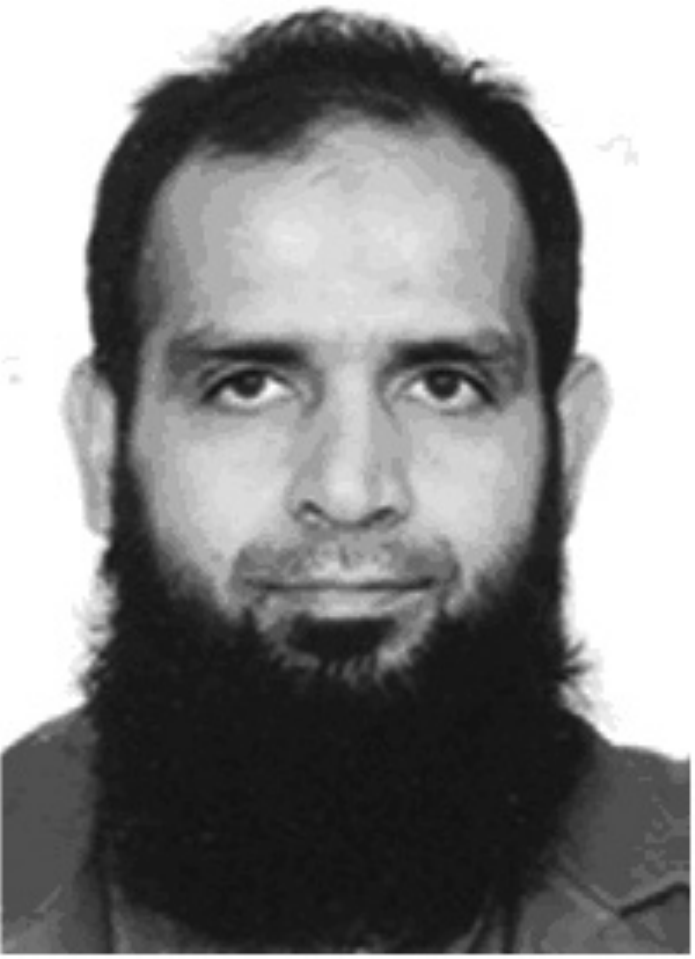}}]{Muddassar Farooq}
completed his D.Sc. in informatics from Technical University of
Dortmund, Germany, in 2006. Currently, he is working as Professor and
Vice Chancellor at Muslim Youth University, Islamabad,
Pakistan. He is also the director of Next Generation Intelligent
Networks Research Center (nexGIN RC). His research interests include
nature inspired applied systems, nature
inspired computing, and network security systems. He has authored several papers and book chapters in these areas.

\end{IEEEbiography}

\end{document}